
\magnification=1200
\hsize=15truecm
\vsize=23truecm
\baselineskip 19 truept
\voffset=-0.5truecm
\parindent=1cm
\overfullrule=0pt
\font\titolo=cmbx10 scaled\magstep2

\def\pmb#1{\leavevmode\setbox0=\hbox{$#1$}\kern-.025em\copy0\kern-\wd0\kern-
.05em\copy0\kern-\wd0\kern-.025em\raise.0433em\box0}

\def\qmb#1{\leavevmode\setbox0=\hbox{$\scriptstyle#1$}\kern-.025em\copy0
\kern-\wd0\kern-.05em\copy0\kern-\wd0\kern-.025em\raise.0433em\box0}

\def\Ai{\hbox{\hbox{${\cal A}$}}\kern-1.9mm{\hbox{${/}$}}}
\def\Vi{\hbox{\hbox{${\cal V}$}}\kern-1.9mm{\hbox{${/}$}}}
\def\Di{\hbox{\hbox{${\cal D}$}}\kern-1.9mm{\hbox{${/}$}}}
\def\lam{\hbox{\hbox{${\lambda}$}}\kern-1.6mm{\hbox{${/}$}}}
\def\D{\hbox{\hbox{${D}$}}\kern-1.9mm{\hbox{${/}$}}}
\def\A1{\hbox{\hbox{${A}$}}\kern-1.8mm{\hbox{${/}$}}}
\def\A{\hbox{\hbox{${\scriptstyle A}$}}\kern-1.8mm{\hbox{${/}$}}}
\def\V{\hbox{\hbox{${V}$}}\kern-1.9mm{\hbox{${/}$}}}
\def\Parz{\hbox{\hbox{${\partial}$}}\kern-1.7mm{\hbox{${/}$}}}
\def\parz{\hbox{\hbox{${\scriptstyle\partial}$}}\kern-1.7mm{\hbox{${/}$}}}
\def\B{\hbox{\hbox{${B}$}}\kern-1.7mm{\hbox{${/}$}}}
\def\R{\hbox{\hbox{${R}$}}\kern-1.7mm{\hbox{${/}$}}}
\def\si{\hbox{\hbox{${\xi}$}}\kern-1.7mm{\hbox{${/}$}}}

\null

\hfill{DFPD 95/TH/61}
\hfill{hep-th/9511100}
\vskip 3truecm
\centerline{\titolo BOSONIZATION AND DUALITY}
\centerline{\titolo IN CONDENSED MATTER SYSTEMS}

\vskip 0.5truecm

\centerline{P.A. Marchetti}

\centerline{\it Dipartimento di Fisica, Universit\`a di Padova and}
\centerline{\it INFN, Sezione di Padova}
\centerline{\it I--35131 Padova Italy$^*$}

\vskip 0.5truecm

\midinsert
\baselineskip 15truept
\centerline{\bf Abstract}
\vskip 0.3truecm

\noindent
We show that abelian bosonization of 1+1 dimensional fermion systems
can be interpreted as duality transformation and, as a conseguence,
it can be generalized to arbitrary dimensions in terms of gauge
forms of rank $d-1$, where $d$ is the dimension of the space.
This permit to treat condensed matter systems in $d>1$ as gauge
theories. Furthermore we show that in the ``scaling" limit the
bosonized action is quadratic in a wide class of condensed
matter systems.

\endinsert

\vskip 4truecm

\noindent
Talk given at ``Common trends in Condensed Matter and High
Energy Physics", September 3--10, 1995 -- Chia.

\vskip 1truecm
\midinsert
\baselineskip 16truept
\noindent
$^*$ Supported in part by M.P.I. . This work is carried out in the
framework of the European Community Research Programme
``Gauge Theories, Applied Supersymmetry and Quantum Gravity,
contract SC1--CT92--D789.

\endinsert

\vfill\eject

\noindent
{\bf 1. Introduction}
\vskip 0.3truecm
The procedure of writing $1+1$ dimensional fermionic systems in terms of
boson fields \underbar{(bosonization)} has now a long history [1].
(Few years
ago a somewhat different procedure of bosonization have been discovered
for $2+1$ dimensional systems, involving the introduction of Chern--Simons
gauge fields and generalising Jordan--Wigner transformation [2]; here we are
not dealing with it).

Only recently it has been realized [3,4] that the abelian bosonization of
one--dimensional systems is a special case of a more general (and now
obiquitous!) transformation:\underbar{duality}, \underbar{without}
\underbar{restriction on dimensions}. It then follows that one can
generalize the abelian bosonization to arbitrary dimensions (although in
general it is less powerful) in terms of \underbar{gauge forms}
(antisymmetric tensor gauge fields) of rank $d-1$, where  $d$ is the
space dimension, playing the role of the scalar field in $d=1$.

One can then apply the bosonization in particular to condensed matter
systems [4]. This permits to treat \underbar{non--relativistic Fermi systems}
with positive density at $T \sim 0$ \underbar{as gauge theories} ($d > 1$)
and to apply to them methods developed in the analysis of gauge
theories in high--energy physics. As an application we will briefly discuss
the Wilson criterion for the existence of the charge operator.

Furthermore, \underbar{for a large class of systems}
(free electron gas, insulators,
Hall fluids, B.C.S. superconductors...) one can prove that the
\underbar{bosonic action is quadra\-}\underbar{tic} in a suitably defined
\underbar{``scaling limit"}. It also follows from general properties
of bosonization that \underbar{density--density} or
\underbar{current--current} (two--body) \underbar{perturbations} are
exactly \underbar{gaussian in the bosonic field}, this lead to the
conjecture that it is possible a classification of large--scale charge
properties of condensed matter systems in \underbar{universality classes},
using \underbar{vacuum polarization tensor}.

Some applications of these ideas are sketched
and the relation of this bosoni\-zation procedure with Luther--Haldane
bosonization of Fermi liquids is exhibited.

\vskip 0.5truecm
\noindent
{\bf 2. Bosonization}
\vskip 0.3truecm

Bosonization corresponds roughly to the following statement: in
$d=1$ a quantum theory of a fermion field $\hat \psi$ with linear
dispersion relation can  be written in terms of a quantum scalar field
$\hat \phi$ with quadratic dispersion relation, describing fluctuations of
fermion--antifermion pairs. [In condensed matter systems, fermions with
linear dispersion are obtained linearizing the dispersion relation of
non--relativistic fermions around the two points of the Fermi surface,
a procedure legitimate if we are interested in large scale properties.
In high--energy physics $\hat \psi$ is just the massless Dirac field].
More precisely, setting the Fermi velocity $v_F=1$ in condensed matter
systems, and the velocity of light $c=1$ in relativistic systems, let
$\psi, \bar \psi$ denote two--component Grassman fields and $\phi$ a complex
field describing in the euclidean path--integral formalism a massless Dirac
field $\hat \psi$ and a neutral scalar field $\hat \phi$, respectively.
Bosonization can be stated as follows: the (euclidean) correlation
functions corresponding to the lagrangian ${\cal L}_F = \bar \psi \Parz
\psi$ of the (euclidean) fields $:\bar \psi \gamma_\mu \psi:(x)
\equiv {\cal
J}_\mu (x)$ (the 2--current), $:\bar \psi \psi:(x), \psi_R (x) \equiv ({1
+ \gamma_5 \over 2}) \psi(x)$, $\psi_L (x) \equiv ({1 - \gamma_5 \over
2}),$ are identical to the correlation
functions corresponding to the lagrangian ${\cal L}_B = {1 \over 8 \pi}
(\partial_\mu \phi)^2$ of the fields ${1 \over 2\pi} \varepsilon_{\mu\nu}
\partial^\nu \phi (x)$, $:\cos \phi:(x)$, $: e^{+{i \over 2}
\phi (x)}: D(x,1)$,$:e^{- {i \over 2} \phi (x)}: D(x,1)$,
.... where : \ : denotes normal ordering (and from now on it will be
omitted) and $D(x,1)$ is a disorder field [4,5] creating a vortex of unit
vorticity at $x \in {\bf R}^2$.

It has been realized in [3] (and independently in a preliminary version of
[4]) that this bosonization formulas are just a special version of the
\underbar{duality transforma\-}\underbar{tion} in $d=1$.

\vskip 0.5truecm
\noindent
{\bf 3. Duality}
\vskip 0.3truecm

\qquad
We now outline the general structure of duality.

\noindent
\underbar{Remark on
notations} \  To avoid topological complications we work in ${\bf R}^{d+1}$,
furthermore to avoid the cumbersome use of multiindices we use the language
of forms: given an antisymmetric tensor field $F_{\mu_1, .. \mu_k}$ we
define a $k$-form $F \equiv {1 \over k!} F_{\mu_1... \mu_k}
dx^{\mu_1} \wedge ... \wedge dx^{\mu_k}$, where $\wedge$ is the wedge
($\equiv$ antisymmetric tensor) product. We denote by $\Lambda^k ({\bf
R}^{d+1})$ the group of $k$-forms on ${\bf R}^{d+1}$ under pointwise
addition, by $d$ the exterior differential $d: \Lambda^k \rightarrow
\Lambda ^{k +1}$, with $dF = {1 \over k !} \partial_\mu F_{\mu_1 ...
\mu_k} dx^\mu \wedge dx^{\mu_1} \wedge ... \wedge dx^{\mu_k}$, by $^*$
the Hodge--star  $^*: \Lambda^k \rightarrow \Lambda^{d+1 - k}$,
with $^*F= {1 \over k!} {1 \over (d+1 - k)!} F^{\mu_1... \mu_k}
\varepsilon _{\mu_1... \mu_k ... \mu_{d+1}}$ $dx^{\mu_{k +1}}
\wedge ... \wedge dx^{\mu_{d+1}}$ and by (,) the inner product in
$\Lambda^k$: for $F, F' \in \Lambda^k ({\bf R}^{d+1})$

$$
(F, F') = {1 \over k !} \int d^{d+1} x F_{\mu_1 ... \mu_k} F^{\mu_1...
\mu_k} (x) = \int F \wedge^* F'.
$$

\noindent
To discuss duality we need two basic facts

\noindent
i) \underbar{Poincar\`e lemma}: let $F \in \Lambda^k ({\bf
R}^{d+1})$ be closed, i.e. $dF=0$, then there exists $A \in \Lambda^{k
-1} ({\bf R}^{d+1})$ such that $F= dA$

\noindent
ii) Denote by $\Lambda^k / d \Lambda^{k -1}$ the quotient group
of equivalence classes $[F] = \hfill\break \{F' \in \Lambda^k | F' -
F = d \zeta,
\zeta \in \Lambda^{k-1}\}$, then $d$ establishes a group isomorphism
between $\Lambda^k /d \Lambda^{k-1}$ and the image of $d$ in $\Lambda^{k+1}$,
the group of closed ($k+1$)-forms.

\vskip 0.5truecm
\noindent
\underbar{Basic formula}
\vskip 0.3truecm

Suppose we can formally write the euclidean partition function of a quantum
field theory in terms of a $k$-form  $F$ in ${\bf R}^{d+1}$ as

$$
Z = \int {\cal D} F e^ {-S(F)} \delta (dF).  \eqno(1)
$$

\noindent
Then we have a ``dual formulation" of such a theory in terms of a
($d-k$)-form $B$, invariant under the gauge transformation

$$
B \rightarrow B + d \zeta, \zeta \in \Lambda^{d - k -1} ({\bf
R}^{d+1})
$$

\noindent
or, alternatively, in terms of a ($d - k +1$)-form  $H$, satisfying $dH
=0$.

To find, (heuristically) this dual formulation we first express the
constraint $dF=0$ in (1) by a Fourier representation of the
$\delta$--functional:

$$
\delta(dF)= \int {\cal D} [B] e^{i\int F \wedge dB}
$$

\noindent
where ${\cal D} [B]$ denotes the normalized measure on the gauge
equivalence classes

$$
[B] = \{B' \in \Lambda^{d-k} ({\bf R}^{d+1})| B' - B = d \zeta,
\zeta \in \Lambda^{d-k} ({\bf R}^{d+1}) \}.
$$

\noindent
Alternatively one can use the gauge--fixing + Faddev--Popov ghost procedure
to properly define a BRS invariant measure for $B$ [6].
Define $\tilde S (dB)$ through the functional integral Fourier
transform

$$
e^{-\tilde S (dB)} \equiv \int {\cal D} F e^{- S(F)} e^{i \int F \wedge
dB}. \eqno(2)
$$

\noindent
Then

$$
Z= \int {\cal D} F e^{-S(F)} \delta(dF)=
\int {\cal D} F e^{-S(F)} \int {\cal D} [B] e^{i \int F \wedge dB} =
\eqno(3)
$$
$$
\int {\cal D} [B] e^{-\tilde S(dB)} = \int {\cal D} H e^{- \tilde S (H)}
\delta (dH),
$$

\noindent
where in the last equality we used the previously defined properties
i) and ii).

\vskip 0.5truecm
\noindent
\underbar{Examples}: a) \underbar{Abelian gauge theories}

Consider a quantum field theory described in the euclidean formulation in
terms of a ($k-1$)-form $A$ and whose action $S$ is invariant under the
gauge transformation

$$
A \rightarrow A + d \zeta, \zeta \in \Lambda^{k -2}.
$$

\noindent
Then, using the isomorphism established by $d$, one can change variable in
the path--integral representation of the partition function from $A$ to a
$k$-form $F$, constrained by $dF =0$:

$$
Z = \int {\cal D} [A] e^{-S(dA)} = \int {\cal D} F e^{-S(F)} \delta (dF).
$$

\noindent
[For $k =1$, $d\zeta$ is replaced by a closed $0-$form, i.e.
a constant]. The corresponding duality is widely known as
\underbar{Wegner -- t'Hooft duality} [7].
In the lattice version, in $d=1$ for ${\bf Z}_2$-valued $0$-forms,
it has already been introduced by \underbar{Kramers and Wannier} [8] in 1941
for the Ising model.

\vskip 0.5truecm

\qquad
b) \underbar{Theories with global abelian gauge invariance}

\vskip 0.3truecm
Consider a quantum field theory expressed in euclidean formalism in terms
of  ``charged" fields $\chi, \chi^*$ whose action $S(\chi, \chi^*)$ is
invariant under an abelian (e.g. $U(1)$) global gauge transformation

$$
\chi(x) \rightarrow e^{i \alpha} \chi (x), \ \chi^* (x) \rightarrow e^{-i
\alpha} \chi^*(x).
$$

\noindent
We promote the global gauge invariance to a local gauge
invariance introducing a minimal coupling between $\chi, \chi^*$
and a $U(1)$-gauge field $A$. Integrating over $A$ and setting $dA=0$ one
recovers the original theory. In formulas, for the partition function we
have:

$$
Z= \int {\cal D} \chi {\cal D} \chi^* e^{-S(\chi, \chi^*)} = \int {\cal D}
\chi {\cal D} \chi^* {\cal D} A
e^{-S(\chi, \chi^*, A)} \delta(dA) = \int {\cal D} A e^{-S(A)} \delta (dA),
$$

\noindent
where $S(\chi, \chi^*, A)$ is gauge invariant and $S(A)$ is the effective
action obtained integrating out $\chi, \chi^*$. A suitable version of
duality  for models of class b) gives the
\underbar{abelian $T$--duality} [9] and as
we shall see, bosonization is
just duality in case $b$ when $\chi$ is the Fermi field $\psi$.

\vskip 0.5truecm
\noindent
{\bf 4. General features of duality}
\vskip 0.3truecm

\noindent
Let us outline some general properties of duality following simply from the
definition.

\noindent
1) From the property that the square of a Fourier transformation is parity
it follows that:

$$
\tilde {\tilde S} (F) = S (-F)
$$

\noindent
2) \underbar{Correlation functions at non--coinciding arguments of}
$-i({\delta S \over
\delta F})_{\mu_1... \mu_k}$ \underbar{are given} \hfill\break
\underbar{in the dual theory by correlation functions of}
$(^*dB)_{\mu_1 ... \mu_k}$ (or $(^*H)_{\mu_1...\mu_k})$.

In fact, denoting by $\langle \ \rangle$ the expectation value
in the original $(F)$
theory and by $\langle \ \tilde \rangle$ the expectation values in the dual
($B$  or $H$) theory, and omitting all indices, we have that

$$
\langle \prod_j \Bigl(-i {\delta S \over \delta F(x_j)} \Bigr) \rangle =
Z^{-1} \int {\cal D} [B]
\int {\cal D} F \prod_j
\Bigl(- i {\delta \over \delta F(x_j)} \Bigr)
e^{-S(F)} e^{i \int F \wedge dB} =
$$
$$
Z^{-1} \int {\cal D} [B] \int {\cal D} F
e^{-S(F)}
\prod_j \Bigl(-i {\delta \over \delta F(x_j)} \Bigr) e^{i \int F \wedge dB} =
$$
$$
= \langle \prod_j (^* dB) (x_j) \tilde \rangle= \langle \prod_j
(^* H) (x_j) \tilde  \rangle,\eqno(4)
$$

\noindent
where in the second equality integration by parts has been used.

\noindent
For models in class a) the equation of motion of the $F$ theory
are written as
$d^* {\delta S \over \delta F} =0$. They are mapped by duality to the
Bianchi identities $dH=0$ and conversely the Bianchi identities given by
$dF =0$ are mapped to $d^* {\delta \tilde S \over \delta H} =0$.
Hence, duality interchanges equations of motions and Bianchi identities.

\vskip 0.3truecm
\noindent
\underbar{Remark}  In $d=3$, for $k=2$, also $H$
is a two-form and we denote it by $\tilde F$. Under duality

$$
{-i {\delta S \over \delta F} \choose ^*F} \rightarrow {^* \tilde F
\choose i {\tilde {\delta S} \over \delta \tilde F}} =
\pmatrix{0 & 1 \cr
-1 & 0 \cr} {-i {\delta \tilde S \over \delta \tilde F} \choose ^* \tilde
F}. \eqno(5)
$$

\noindent
Furthemore for such values of $d,k$ one can add to the action the
$\theta$ term ${\theta \over 2\pi}$ $\int F \wedge F$ and the theory
is invariant under $\theta \rightarrow \theta + 2\pi$ .
Under this transformation

$$
{-i {\delta S \over \delta F} \choose ^*F} \rightarrow {-i {\delta S
\over \delta F}+ ^*F \choose ^*F}=
\pmatrix{1 & 1 \cr
0 & 1 \cr} {-i {\delta S \over \delta F} \choose ^*F}.\eqno(6)
$$

One recognizes the $2\times 2$ matrices in (5) (6) as the $S$
and $T$ generators
of $SL (2, {\bf Z})$, hence one can construct a full $SL(2,{\bf Z})$
group of equivalent descriptions of the theory. An $N=2$ supersymmetric
version of these transformation is a building block of Seiberg-Witten
discussion of low-energy $N=2$ Super-Yang Mills, with gauge group
$SU(2)$ [10].

\noindent
For models in class b),$-i {\delta S \over \delta A_\mu} (x) =
{\cal J}_\mu (x)$, the current associated to the global $U(1)$ symmetry,
hence current correlation functions are expressed in the dual theory as
$^* dB$--correlation functions and the analogue of the equation of motion
in models of class a) is just current conservation:

$$
d^* {(-i {\delta S \over \delta A})}= d^* {\cal J} =0
$$

\noindent
3) \underbar{order--disorder duality}

\vskip 0.3truecm
Let $\Sigma_p$ be a $p$-dimensional surface and denote
by $\tilde {\Sigma}_p$ its Poincar\`e dual ($d+1-p$)-current, so that
for $F \in \Lambda^p ({\bf R}^{d+1})$ we have:

$$
\int_{\Sigma_p} F  = \int F \wedge \tilde {\Sigma_p}.
$$

\noindent
In a theory of gauge forms $F$ of rank $k$ the ``\underbar{Wilson
loop" order field} $W_\alpha (\Sigma_k)$, $\alpha \in {\bf R}$, is
defined by

$$
W_\alpha (\Sigma_k)= e^{i \alpha \int_{\Sigma_k} F}= e^{i \alpha \int F
\wedge \tilde {\Sigma}_k} \eqno(7)
$$

\noindent
and it measures the ``magnetic flux" through $\Sigma_k$.

The \underbar{``Wegner-- t'Hooft" disorder field} $D_\alpha
(\Sigma_{d+1-k})$ in the same theory is obtained instead shifting $F$ in
the action by $\alpha \tilde {\Sigma}_{d+1 - k}$, i.e.

$$
\langle D_\alpha \Bigl(\Sigma_{d+1-k} \Bigr) \rangle = \langle
e^{-[S(F - \alpha \tilde
{\Sigma}_{d+1-k}) - S (F)]} \rangle \eqno(8)
$$

\noindent
and it measures the ``electric flux" through $\Sigma_{d+1-k}$
(Normalisation factors are omitted in (7) (8), see [4]).

\noindent
\underbar{Duality exchanges Wegner -- t'Hooft disorder field and Wilson
loop order field}, in fact

$$
\langle W_\alpha (\Sigma_k) \rangle = \int {\cal D} F {\cal D} [B] e^{-S(F)}
e^{i \int F \wedge dB} e^{i \alpha \int F \wedge \tilde {\Sigma}_k}
$$
$$
= \int {\cal D} F {\cal D} [B] e^{-S(F - \alpha \tilde {\Sigma}_k)} e^{i \int F
\wedge dB} = \langle D_\alpha (\Sigma_k) \tilde \rangle,
$$

\noindent
where in the second equality we use the change of variable $F \rightarrow F
+ \alpha \tilde {\Sigma}_k$.

\vskip 0.5truecm
\noindent
{\bf 5. Bosonization in condensed matter system}

\vskip 0.3truecm
It has been proved in [3,4], that abelian bosonization is
duality for a model in class b) with $\chi
\equiv \psi$ the massles Dirac field in $d=1$.

\noindent
The proof for the partition function is immediate [3] using the old result
by Schwinger

$$
\int {\cal D} \bar \psi {\cal D} \psi e^{-\int \bar \psi ({\parz} - {\A})\psi}
= e^{- {1 \over 2\pi} (dA, \Delta^{-1} dA)}
$$

\noindent
where $\Delta$ is the two--dimensional laplacian. In fact,
with $B \in \Lambda^0 ({\bf R}^2)$,

$$
Z = \int {\cal D} \bar \psi {\cal D} \psi e^{-\int \bar \psi {\parz} \psi}=
\int {\cal D} \bar \psi {\cal D} \psi {\cal D} [A] {\cal D} B
e^{- \int \bar \psi ({\parz} - {\A}) \psi}
e^{i \int A \wedge dB} =
$$
$$
\int {\cal D} [A] e^{-{1\over 2\pi} (dA,
\Delta^{-1} dA)} e^{i \int A \wedge dB}
= \int {\cal D} B e^{-{\pi \over 2} (B,\Delta B)} = \int {\cal D} \phi
e^{-{1 \over 8\pi} (\partial_\mu \phi)^2},
$$

\noindent
where we identify $B \equiv {\phi \over 2\pi}$.
The proof for current correlation functions [3] follows from property 2)
in sect 4 at non--coinciding arguments and can be extended also to
coinciding points, using gauge invariance [4]. The proof for fermion
correlation functions is slightly more involved, see [4].

A basic message we learn from this identification
is the possibility to extend bosonization to arbitrary Fermi systems
replacing $\phi$ by a ($d-1$)-gauge form $B$ and in particular one can obtain
a bosonized (dual) action $\tilde S(dB)$ for condensed matter systems in
arbitrary dimensions.

However,
the problem we are faced on, is that even if bosonization as
duality is always in principle applicable, it becomes useful only if
$\tilde S (dB)$ has a tractable form at least for some ``reference
systems". This is not true in general, of course; in this respect
Schwinger result for massless Dirac fields in $d=1$ is very special!
However, one can hope that $\tilde S (dB)$ simplify at large scales. To
discuss large-scale properties of $T \sim 0$ systems we proceed as follows:
we confine our Fermi system in cubes $\Omega_\lambda = \{\lambda x| x \in
\Omega \}, \Omega$ being a fixed cube in ${\bf R}^d$ and $\lambda,
1 \leq \lambda <
\infty$, a scale parameter. We keeps the particle density constant
and couple the fermions to a $U(1)$-gauge field
$A^\lambda (\lambda x) \equiv \lambda^{-1} A(x)$
where $A$ is an arbitrary but $\lambda$-independent gauge potential.
Let $S^{\Omega_\lambda} (A^\lambda)$ denote the corresponding
gauge-invariant action and expand it in Laurent series around $\lambda =
\infty$:

$$
S^{\Omega_\lambda} (A^\lambda)_{\sim\atop\lambda\nearrow\infty}
\lambda^k \sum^\infty_{n=0} \lambda^{-n} S^{(n)} (A). \eqno(9)
$$

We call the leading term in this expansion the \underbar{``scaling limit"}
of the
effective action $S(A)$ of our system and we denote it by $S^\star (A)$.
It is expected to give a good description of large scale properties of
$S(A)$. The dual action is denoted by $\tilde S^\star (dB)$. Somewhat
remarkably, one can prove [4,11,12] that $S^\star (A)$,
and hence $S^\star (dB)$,
is quadratic for insulators (I), Hall fluids (H), free
electron gas (F), B.C.S. superconductors (S). [The proof does not
use small - $A$ arguments nor in general follows from dimensional
analysis, furthermore an analogous statement is false for an
analogous treatment of the spin degrees of freedom, where $A$ is
non--abelian.
Let us outline the basic ideas of the proof in cases I,H,F.]
The proof is easy if the spectrum is gapful (I,H). In fact, as a
consequence,
the connected current correlation functions $\langle \prod_j
{\cal J}^{\mu_j} (x_j)
\rangle^c$ decay exponentially as $|x_i - x_j| \rightarrow \infty$, so that in
the scaling limit they become distributions with point-like support,
given by linear combinations of
$\delta$-functions and a finite number of derivative
of $\delta$. In turn, one easily realizes that

$$
\prod_j {\delta \over \delta A_{\mu_j} (x_j)} S(A) = \langle \prod_j {\cal
J}^{\mu_j} (x_j) \rangle^c,
$$

\noindent
so that connected current correlation functions are just the coefficients
of a series expansion of $S(A)$ in power of $A$. As a result  $S^\star(A)$ is
local and its form can then be determined by using dimensional arguments
and symmetries.

\noindent
For example for parity preserving \underbar{rotation symmetric insulators}
one finds [11]

$$
S^\star (A) = \int d^{d+1} x \{c_1 F_{ij} (A) F^{ij} (A) + c_2 F_{0i} (A)
F^{0i} (A) \}(x) \eqno(10)
$$

\noindent
where $F^{\mu\nu} (A) = \partial^{[\mu} A^{\nu ]}$, so that $S^\star(A)$ is
Maxwell--like; for \underbar{Laughlin fluids} (Hall fluids at Laughlin
plateaux, where only a $U(1)$ symmetry appears) one finds, as a conseguence
of broken parity [11] the Chern-Simons action

$$
S^\star (A) = c \int A \wedge dA. \eqno(11)
$$

The proof [4,12] is less easy for electron gas and superconductors where the
absence of gap forbids any argument of locality. Let us outline the
idea for the electron gas, it will turn out that the result follows,
roughly speaking, treating a $d$-dimensional Fermi surface as the union of
1-dimensional Fermi surfaces corresponding to its rays.
We start noticing that at large scale only regions close to the Fermi
surface contribute to the fermion propagator, which can be approximately
written as

$$
\langle \psi^* (\lambda x) \psi (\lambda y) \rangle_{\sim\atop\lambda
\nearrow
\infty} {1 \over \lambda} \int_{S^{d-1}} d {\pmb \omega} e^{i k_F
\lambda {\qmb \omega} \cdot ({\bf x} - {\bf y})} ({k_F \over 2\pi})^{d-1}
G_{\qmb \omega} \Bigl(x_0 - y_0, {\pmb \omega} \cdot ({\bf x} \cdot {\bf
y}) \Bigr), \eqno(12)
$$

\noindent
with

$$
G_{\qmb \omega} (x_0, {\pmb \omega} \cdot {\bf x}) = \int {d k_0 \over 2\pi} {
dk_1 \over 2\pi} {e^{-i(k_0 x_0 + k_1 {\qmb \omega} \cdot {\bf x})} \over ik_0
- v_F k_1} \eqno(13)
$$

\noindent
where $d {\pmb \omega}$ is the uniform measure on the
$d-1$--dimensional unit sphere $S^{d-1}, k_F$ is the Fermi momentum
and $v_F$ the Fermi velocity and from now on we set $v_F =1$. Let us
introduce a field $\psi_{\qmb \omega}$ for each point indexed by ${\pmb
\omega}$ of the Fermi surface and using a ``relativistic notation" set
$\psi_{[{\qmb \omega}]}$ = $\psi_{\qmb \omega} \choose \psi_{-{\qmb\omega}}$,
where $[{\pmb \omega}] \equiv \{{\pmb \omega}, - {\pmb \omega}\}$.
Then, the approximate formula (12) is recovered identifying

$$
\psi (\lambda x)_{\sim\atop\lambda\nearrow \infty} \int d {\pmb \omega}
e^{-i k_F \lambda \ {\qmb \omega} \cdot {\bf x}} \psi_{{\qmb \omega}}
(\lambda x^0, \lambda \ {\pmb \omega} \cdot {\bf x})
$$

\noindent
and replacing the free electron action in the
scaling limit by the integral of one--dimensional massless Dirac action:

$$
({k_F \over 2\pi})^{d-1} \int_{S^{d-1}} d [{\pmb \omega}] \int d^{d+1}
x \bar \psi_{[{\qmb \omega}]} \Parz_{\qmb \omega} \psi_{[{\qmb \omega}]}
(x)\equiv S_0
(\psi_{[{\qmb \omega}]}, \bar \psi_{[{\qmb \omega}]}) \eqno(14)
$$

\noindent
where $\partial^\mu_{\qmb \omega} = (\partial_0, {\pmb \omega} \cdot
{\pmb \nabla})$.
The possibility of expressing the action in the limit
of $\lambda \nearrow \infty$ as an integral over one-dimensional actions
persists if we couple the free fermions to a gauge field $A$, in fact

$$
S(\psi, \psi^*, A^\lambda)_{\sim\atop\lambda \nearrow
\infty}
S_0 (\psi_{[{\qmb \omega}]}, \bar \psi_{[{\qmb \omega}]})
+ \lambda^d i \int d [{\pmb \omega}] \int d^{d+1} x \  A_\mu^{\qmb
\omega} (x)
j^\mu_{[{\qmb \omega}]} (x; \lambda) \eqno(15)
$$

\noindent
where
$A^{\qmb \omega}_\mu = (A_0, {\pmb \omega} \cdot {\bf A})$ and

$$
j^\mu_{[{\qmb \omega}]} (x; \lambda) = {1 \over \lambda} \int d [{\pmb
\omega'}]
e^{-i \lambda k_F ({\qmb \omega} - {\qmb \omega'})\cdot {\bf x}} \bar
\psi_{[{\qmb \omega '}]} (x^0, {\bf x} \cdot {\pmb \omega'}) \gamma^\mu
\psi_{[{\qmb \omega}]} (x^0, {\bf x} \cdot {\pmb \omega}).
$$

\noindent
\underbar{Remark} \ Formally, in the limit $\lambda \nearrow \infty$

$$
j_{[{\qmb \omega}]}^\mu  (x; \lambda) \rightarrow  {1 \over \lambda^d} \delta
({\bf x} \wedge {\pmb \omega}) \bar \psi_{[{\qmb \omega}]}\gamma_\mu
\psi_{[{\qmb \omega}]} (x^0, {\bf x} \cdot {\pmb \omega}),
$$

\noindent
however perturbation by $A$ and $\lambda \nearrow \infty$ limit do
\underbar{not} commute! [12].

Since for every ray $[{\pmb \omega}]$ in (15), the action
of $\psi_{[{\qmb \omega}]}$ is
1-dimensional, the effective action is quadratic in $A^0$ and as a
consequence the full effective action is also quadratic in $A$, being
integral of quadratic actions.
One can easily verify that denoting by $(\Pi^\star_F)^{\mu\nu}$
the scaling limit
of the free electron vacuum polarization tensor,

$$
S^\star (A) = \int d^{d+1} x d^{d+1} y A_{\mu} (x)
(\Pi^\star_F)^{\mu\nu} (x-y) A (y) \equiv (A, \Pi_F^\star A). \eqno(16)
$$

\noindent
\underbar{Remark} \ \underbar{Relation with Luther--Haldane
bosonization} [12]

\smallskip

Since $\psi_{[{\qmb \omega}]}$ is a 1+1 Dirac massless Fermi field, one can
directly bosonize (15) in terms of a scalar real field
$\varphi_{[{\qmb \omega}]}$, and one obtains

$$
j^\mu_{[{\qmb \omega}]} = {1 \over 2\pi} \epsilon^{\mu\nu} \partial_{{\qmb
\omega} \nu} \varphi_{[{\qmb \omega}]}
$$

\noindent
This procedure gives the (euclidean version of the) Luther-Haldane
bosonization [13]. The relation of $\varphi_{[{\qmb \omega}]}$
with the dual field $B$ is given by

$$
{\cal J}^0 = (^*dB)^0 = {1 \over 2\pi} \int d [{\pmb \omega}] {\pmb \omega}
\cdot {\pmb \nabla} \varphi_{[{\qmb \omega}]}
$$
$$
{\cal J}^k = (^*dB)^k = {1\over 2\pi} \int d [{\pmb \omega}] \omega^k
\partial_0 \varphi_{[{\qmb \omega}]}
$$

\vfill\eject

\noindent
{\bf 6. \ Adding perturbations}
\vskip 0.3truecm

\noindent
According to property 2) of duality, density-density or
current-current perturbations in the dual theory are quadratic in $B$.
Hence, if we have a ``reference" system with scaling
limit effective action $S^\star_0 (A)= (A, \Pi^\star A)$ and,
as a conseguence,
bosonized action $S^\star_0 (dB)= (^* dB,(\Pi^\star)^{-1}$ $^* dB)$,
one would be tempted to say that the
perturbed system have a quadratic (!) scaling limit bosonized action given by
$S^\star (dB) = (^* dB, ((\Pi^\star)^{-1} +V^\star) ^* dB)$, where $V$
is the perturbation kernel.

However this holds only if the following \underbar{perturbative
assumption} $(P)$ is satisfied: perturbation and scaling limit commute.

\noindent
\underbar{Remark}: What could happen is that the perturbation drive the
reference system away from its fixed point in the scaling limit. A typical
example is obtained choosing $S_0$ as the action of free fermions
and $V$ a Cooper
interaction: the scaling limit of the perturbed system is known to describe
a superconductor!.
Assumption $P$ can be argued to hold for perturbed free systems if $V$ is
long range and the Cooper channel is tunnel off [12,14].

If assumption $P$ holds, then, in the scaling
limit of the perturbed system, the two--point current correlation function
is given by

$$
\langle {\cal J}^\mu (x) {\cal J}^\nu (y) \rangle^\star = \langle(^*dB)^\mu (x)
(^*dB)^\nu (y) \rangle^\star
= [((\Pi^{\star})^{-1} + V^\star)^{-1}]^{\mu\nu} (x,y), \eqno(17)
$$

\noindent
Equation (17) is exactly the result
of R.P.A.! Hence, assumption $P$ implies exactness of R.P.A. in the scaling
limit. This explains e.g. why in a free electron system perturbed by a
Coulomb potential
the plasmon gap obtained by R.P.A. coincides with
the non--perturbative exact result obtained by Morchio and Strocchi [15] by
a ``generalized Goldstone theorem".

To summarize, bosonization combined with assumtpion $P$ gives a method to
treat non--relativistic $T \sim 0$ systems in the scaling limit as gauge
theories for $d > 1$. One can then apply to them the techniques elaborated in
the analysis of gauge theories. As an application we discuss the Wilson
criterion for the existence of the charge operator.

\vskip 0.5truecm
\noindent
{\bf 7. Existence of the charge operator}
\vskip 0.3truecm

As remarked before, by duality a Wilson loop $W_\alpha (\Sigma_d)$ measures the
charge contained in a $d$-dimensional surface $\Sigma_d$ in the dual $(B)$
theory.
One can prove that, if it exists, the charge operator ${\cal Q}$ can be
defined through the (weak) limit

$$
e^{i \alpha {\cal Q}} = \lim_{R \nearrow \infty}
{W_\alpha (\Sigma^R_d) \over \langle W_\alpha (\Sigma^R_d) \tilde \rangle},
\eqno(18)
$$

\noindent
where $\Sigma^R_d$ is a ball of radius $R$ in the time $0$ (hyper-)plane. The
normalization ensures that if ${\cal Q}$ exists it annihilates the vacuum.
For the existence of the limit (18), one can use the Wilson criterion,
proved to be correct for many lattice gauge theories:
the limit exists if for $R \nearrow \infty$

$$
\langle W_\alpha (\Sigma^R_d) \rangle \geq e^{- c|\partial \Sigma^R_d|},
$$

\noindent
where $|\partial \Sigma|$ denote the volume of the boundary of
$\Sigma$, i.e. if the Wilson loop has ``perimeter decay", and the limit
does not exist if it has a faster decay, e.g. as $R \nearrow \infty$

$$
\langle W_\alpha (\Sigma^R_d) \rangle \leq
e^{-c |\partial \Sigma^R_d| {\rm ln} R}.
$$

\noindent
In the $B$--theory one can easily compute

$$
\langle W_\alpha (\Sigma^R_d) \rangle_{\sim \atop R \nearrow \infty}
\cases{e^{-c \left|\partial \Sigma^R_d \right|} & {\rm I} \cr
1 & H \cr
e^{-c \left|\partial \Sigma^R_d \right| {\rm ln} R} & {\rm F} \cr
e^{-c \left|\partial \Sigma^R_d \right| {\rm ln}  R} & {\rm S}. \cr}
$$

\vskip 0.3truecm
This implies existence of the charge operator ${\cal Q}$ for insulators and
Hall fluids, so that in these systems ${\cal Q}$ defines a
superselection rule. Viceversa, ${\cal Q}$ does not exists for the free
electron gas and for superconductors, signalizing
that charge fluctuations diverge in the thermodynamic limit.
Under assumption $P$ it follows that
for a long-range repulsive density-density perturbation we obtain
perimeter decay also for perturbed free systems and superconductors: the
long range perturbation depresses charge fluctuations and ${\cal Q}$ is
again well defined.

\vfill\eject
\null
\vskip 0.8truecm

{\bf References}

\noindent
\item{[1]} E. Lieb, D. Mattis, J. Math. Phys. {\bf 6} (1965), 304;
S. Coleman, Phys. Rev. {\bf D11} (1975) 2088;
A. Luther, I. Peschel, Phys. Rev. {\bf B12} (1975), 3908;
S. Mandelstam, Phys. Rev. {\bf D11} (1975), 3026.

\noindent
\item{[2]} G.W. Semenoff, Phys. Rev. Lett. {\bf 67} (1988), 517;
A.M. Polyakov, Mod. Phys. Lett. {\bf A3} (1988), 325;
E. Fradkin, Phys. Lett. {\bf 63} (1989), 322;
M. L\"uscher, Nucl. Phys. {\bf B326} (1989), 557;
J. Fr\"ohlich, T. Kerler, P.A. Marchetti, Nucl. Phys. {\bf B374}
(1992), 511.

\noindent
\item{[3]} C.P. Burgess, F. Quevedo, Nucl. Phys. {\bf B421} (1994), 373;
C.P. Burgess, C.A. Lutken, F. Quevedo, Phys. Lett. {\bf B336} (1994),
18.

\noindent
\item{[4]} J. Fr\"ohlich, R. G\"otschmann, P.A. Marchetti, J. Phys. {\bf A28}
(1995), 1169.

\noindent
\item{[5]} L. P. Kadanoff, H. Ceva, Phys. Rev. {\bf B3} (1971) 3918;
E.C. Marino, J.A. Swieca, Nucl. Phys. {\bf B170} [FS1] (1980), 175;
E.C. Marino, B. Schroer, J.A. Swieca, Nucl. Phys. {\bf B200} [FS 4],
(1982), 473;
J. Fr\"ohlich, P.A. Marchetti, Commun. Math. Phys. {\bf 112} (1987),
343; {\bf 116} (1988), 127.

\noindent
\item{[6]} P.A. Marchetti, M. Tonin, Nuovo Cimento {\bf 63A} (1981), 459 and
references therein;
Y.N. Obukhov, Theor. Math. Phys. {\bf 50} (1982), 229.

\noindent
\item{[7]} F. Wegner, J. Math. Phys. {\bf 12} (1971), 2259;
G. t'Hooft, Nucl. Phys. {\bf B138} (1978), 1.

\noindent
\item{[8]} H.A. Kramers, G. H. Wannier, Phys. Rev. {\bf 60} (1941), 252.

\noindent
\item{[9]} for a review see: E. Alvarez, L. Alvarez Gaum\`e, Y. Lozano ``An
Introduction to T--Duality in string theory"  hepth/9410237.

\noindent
\item{[10]} N. Seiberg, E. Witten, Nucl. Phys. {\bf B246} (1994), 19

\noindent
\item{[11]} J. Fr\"ohlich, U.M. Studer, Rev. Mod. Phys. {\bf 65} (1993), 733

\noindent
\item{[12]} J. Fr\"ohlich, R. G\"otschmann, P.A. Marchetti,
Commun Math. Phys.{\bf 173} (1995), 417.

\noindent
\item{[13]} A. Luther, Phys. Rev. {\bf B19} (1979), 320;
F.D.M. Haldane, Lectures at Varenna 1992, Helv. Phys. Acta {\bf 65}
(1992), 152.

\noindent
\item{[14]} J. Fr\"ohlich, R. G\"otschmann to appear.

\noindent
\item{[15]} G. Morchio, F. Strocchi, Ann. Phys. {\bf 170} (1986), 310.

\bye